\documentstyle[aps,twocolumn,prl]{revtex}

\begin{document}
\draft
\preprint{}
\title{Test for Pairing Symmetry Based on Spin Fluctuations in Organic Superconductors $\kappa$-(BEDT-TTF)$_{2}$X}

\author{
Jian-Xin Li}
\address{
$^{1}$ National Laboratory of Solid States of Microstructure and Department of Physics, Nanjing University, Nanjing 210093, China\\
$^{2}$ The Interdisciplinary Center of Theoretical Studies, Chinese Academy of Sciences, Beijing 100080, China
 }
\maketitle

\begin{abstract}
We propose that the superconducting pairing symmetry of organic superconductors $\kappa$-(BEDT-TTF)$_{2}$X
can be determined by measuring the position in momentum space of the incommensurate peaks of the spin susceptibility. We show that the positions of these peaks are uniquely determined by the symmetry of the superconducting state and the geometry 
of Fermi surface. Based on the effective Hubbard model on an anisotropic triangular lattice, we demonstrate the different incommensurate patters of spin responses for $d_{x^{2}-y^{2}}$- and $d_{xy}$-like pairing states. In addition, we predict the nonexistence of the spin resonance mode in this compound, which is an important feature in high-$T_{c}$ superconductors.  

\end{abstract}

Layered organic superconductors based on the (BEDT-TTF) molecule are rich examples of strongly 
correlated electron systems in low dimensions~\cite{rev}.  Among them, $\kappa$-(BEDT-TTF)$_{2}$X($\kappa$-(ET)) 
has attracted considerable attention, because they share many similarities to cuprate 
superconductors~\cite{kan}, such as the proximity of its superconducting(SC) phase with the 
antiferromagnetic(AF) insulating phase which suggests that it may be an unconventional superconductor. Understanding its SC mechanism could therefore provide important insight into the origin of the unconventional superconductivity. As an essential step for this process, intensive experimental and theoretical investigations have been carried out to identify its SC pairing symmetry, but no consensus has been achieved yet~\cite{sin}. The NMR~\cite{may} and millimeter-wave transmission experiment~\cite{sch} suggests $d_{x^{2}-y^{2}}$ symmetry (A different interpretation has been proposed for the
latter result~\cite{hill}). However, the recent thermal conductivity~\cite{iza} and scan tunneling spectroscopy measurements~\cite{ara} predict that it is more like $d_{xy}$ than $d_{x^{2}-y^{2}}$. Moreover, 
a nodeless $s$-wave SC state is also suggested in recent specific-heat measurements~\cite{els}.
Theoretically, the $d_{x^{2}-y^{2}}$-like pairing symmetry is favored in the model in which the superconductivity
is mediated by strong AF fluctuations~\cite{schmal,kon} and also in a model of small-$q$ phonon mediated superconductivity~\cite{var}. However, it is shown that the $d_{xy}$-like superconductivity is also likely and can dominate over $d_{x^{2}-y^{2}}$ wave, when a more realistic model is considered~\cite{kuroki}. 

In this letter, we propose an orientational-dependent method aiming at a direct determination of the gap anisotropy
in $\kappa$-(ET). It is to measure the $q$-space position of the incommensurate(IC) peaks of the dynamical spin susceptibility by inelastic neutron scatterings. Our proposal is based on: i) it is believed that AF spin fluctuations are strong in this material\cite{rev,kan,may}; ii) the central issue at problem about the gap symmetry is what the direction of the gap 
nodes is or if there are gap nodes. The underlying physics is that the particle-hole excitations, which contribute to the spin susceptibility, are only available for the node-to-node transitions in the SC state, if the excitation energy is lower than 2$\Delta_{0}$($\Delta_{0}$ the magnitude of the SC gap). Therefore, the $q$-space position of spin responses is uniquely determined by the node position and the geometry of the normal state Fermi surface. The Fermi surface in $\kappa$-(ET) has been determined by various experiments~\cite{sin}, so the node position can be reliably determined by measuring the $q$-space structure of spin excitations. This method may be useful specially for $\kappa$-(ET), because the nodes of the $d_{x^{2}-y^{2}}$-wave pairing are different from those of $d_{xy}$-wave pairing by rotating the former $\pi/4$ along the $k_{z}$ axis.

The above argument is illustrated by a numerical calculation based on the Hubbard model on an anisotropic triangular lattice at half filling, which is adopted as the simplest model for $\kappa$-(ET)~\cite{rev,kino}. We find that the spin response is incommensurate in the normal state due to the nesting of Fermi surface. In the SC state, the IC peak appears at $(0.9\pi, 0.9\pi)$ for $d_{x^{2}-y^{2}}$ wave, but around  $(0.35\pi, 0.35\pi)$ for $d_{xy}$ wave. For an isotropic $s$-wave, the peak is suppressed highly and its structure is quite similar to that in the normal state. 
We also investigate the frequency dependence of dynamical susceptibility at the AF wavevector ${\bf Q}=(\pi,\pi)$ and no resonance peak is found in the reasonable range of parameters. It contrases with the case in high-$T_{c}$ cuprates, where a resonance peak is observed at ${\bf Q}$ for YBaCuO, BiSrCaCuO and TlBaCuO~\cite{he}.  

The model Hamiltonian is~\cite{kino},
\begin{equation}
H=-t\sum_{<ij>,\sigma}c_{i,\sigma}^{\dagger}c_{j,\sigma}-t^{\prime}\sum_{<ij>',\sigma}c_{i,\sigma}^{\dagger}c_{j,\sigma}
+U\sum_{i}n_{i,\uparrow}n_{i,\downarrow},
\end{equation}
where the notation is standard except that $<ij>^{\prime }$ denotes the next-nearest-neighbour bond only in one direction. 
Its dispersion is $\epsilon_{k}=-2t(\cos k_{x}+\cos k_{y})-2t^{\prime}\cos (k_{x}+k_{y})-\mu$.
According to the band structure calculations~\cite{fort}, we take $t^{\prime}/t\approx 0.7$ and $U \approx 7t$. The chemical potential $\mu=0.7t$ is determined by fixing the electron density to be at half filling. The Fermi surface is shown in Fig.1. 

In the SC state, the bare spin susceptibility $\chi_{0}({\bf q},\omega)=\chi_{0}({\bf q},i\omega_{n}->\omega+i\delta)$ is given by,
\begin{eqnarray}
\chi_{0}({\bf q},i\omega_{n})&=&-T\sum_{{\bf k},m}G({\bf k},i\omega_{m})G({\bf k+q},i\omega_{n}+i\omega_{m}) \nonumber \\
&+& F({\bf k},i\omega_{m})F({\bf k+q},i\omega_{n}+i\omega_{m}),
\end{eqnarray}
where $G$ and $F$ are the normal and anomalous Green's functions,
\begin{equation}
G={i\omega_{n}+\epsilon_{k} \over (i\omega_{n})^{2}-\epsilon_{k}^{2}-\Delta_{k}^{2}}, 
F={\Delta_{k} \over (i\omega_{n})^{2}-\epsilon_{k}^{2}-\Delta_{k}^{2}},
\end{equation}
with $\Delta_{k}$ the SC pairing order parameter. According to the recent tunneling experiment~\cite{ara},
the maximum gap value $\Delta_{0}$ is taken to be $3.0m$eV=0.04$t$. 

Bulut {\it et al.},~\cite{bulut} have shown that a random-phase (RPA) approximation for the spin susceptibility with a renormalized(reduced) Coulomb coupling $\bar{U}=2t$ gives a fit to the Monte Carlo data for a 2D Hubbard model of $U=4t$. Thus, we consider the renormalized spin susceptibility in the RPA approximation which is given by,
\begin{equation}
\chi(q,\omega)={\chi_{0}(q,\omega)(1-\bar{U} \chi_{0}(q,\omega)})^{-1}
\end{equation}  
where the renormalized vertex $\bar{U}$ instead of the bare $U$ has been used. The system will be instability to an AF N\'eel state if the bare $U$ is used, which is not consistent with the metallic normal state considered here. This is due to the artifact of the RPA approximation, which overestimate the AF correlation as was found in the cases of the cuprate system~\cite{li} and also organic materials~\cite{louati}. Considering the conclusion of 
Bulut {\it et al.}~\cite{bulut} and that the instability to an AF N\'eel state
occurs when $U\geq 3.4t$ , we choose $\bar{U}=3t$ here.   

Four different symmetries of the pairing state are investigated here: i) $\Delta_{k}=\Delta_{0}(\cos k_{x}-\cos k_{y})/2$(pure $d_{x^{2}-y^{2}}$ wave); ii) $\Delta_{k}=\Delta_{0}\sin k_{x}\sin {k_{y}}$(pure $d_{xy}$ wave);
iii) $\Delta_{k}=0.4\Delta_{0}(\cos k_{x}-\cos k_{y})+0.2\Delta_{0}$(mixed $d_{x^{2}-y^{2}}$ wave);
and iv) $\Delta_{k}=0.8\Delta_{0}\sin k_{x}\sin {k_{y}}\pm 0.2\Delta_{0}$(mixed $d_{xy}$ wave). We note that one should differentiate $+0.2\Delta_{0}$ and $-0.2\Delta_{0}$ in case iv) due to the anisotropy of the Fermi surface. This will be discussed below.

Firstly, the momentum(q) dependences of the spin susceptibility in the normal state at $T=0.01t$ are presented in Fig.2(a) and (b) for $\omega=0.02t$ and $\omega=0.2t$, respectively. An obvious and also common feature for both cases
is that the spin response is incommensurate. The IC peaks form two parallel walls near ${\bf Q}=(\pi,\pi)$. 
They come from the excitations with the transition wavevector 1 connecting the nesting pieces of the Fermi surface as shown
in Fig.1.

The spin response in the SC state with $d_{x^{2}-y^{2}}$ symmetry for $\omega=0.02t$ is shown in Fig.3(a). The quasiparticle spectrum now has gap except the four nodes $(\pm k_{0},\pm k_{0})$ along the diagonal direction of the Brillouin zone. For $\omega=0.02t$, which is lower than the maximum SC gap $\Delta_{0}=0.04t$, the excitations away from the nodes are suppressed and only those from nodes to nodes survive. As a result, one can only observe two sharp peak at $(\pi \pm 0.1\pi ,\pi \pm 0.1\pi)$(The $(-k_{0},k_{0})$ to $(k_{0},-k_{0})$ excitations do not give rise to strong peaks due to the lack of
the nesting of Fermi surface along that direction). We have also calculated the result for $\omega=0.2t$ and
find that it is nearly the same as that in the normal state, because the quasiparticle excitations are nearly not affected by the opening of the SC gap. 

We show the results with $d_{xy}$-wave symmetry in Fig.3(b) for $\omega=0.02t$. 
Compared with the result of $d_{x^{2}-y^{2}}$ wave, an obvious difference is that a new peak appears near $(0.35\pi,0.35\pi)$. Moreover, its intensity is stronger than the original one near $(\pi,\pi)$. The origin of this peak can be seen from the Fermi surface shown in Fig.1. For $d_{xy}$-wave symmetry, the four nodes are now at $(\pm p_{1},0),(0,\pm p_{1})$, where $p_{1}$ is the cross point of the Fermi surface with the $k_{x}$ or $k_{y}$ axis. So, the dominating particle-hole excitations at low frequencies are the 
excitations corresponding to the transition 2 as shown in Fig.1. The reason that this peak is along the diagonal direction
instead of the $k_{x}$ or $k_{y}$ direction is that the transition wavevector connecting the nesting piece of the Fermi
surface is along that direction. Now, let us discuss the conditions which this new peak can be observed in experiments, i.e., which its intensity is stronger than the IC peaks around $(\pi,\pi)$. Firstly, the excitation energy for the spin response should be two times smaller than the magnitude of the SC gap $\Delta_{0}$. Otherwise,  
the excitations(transition 1) across the SC gap will overwhelm the node-to-node excitations(transition 2) due to the much stronger nesting effect of Fermi surface near $(\pi/2,\pi/2)$. We show the imaginary part of the dynamical susceptibility along the diagonal direction for different frequencies $\omega=0.02t, 0.08t$ and $0.2t$ in Fig.3(c). When $\omega$ is equal to $2\Delta_{0}=0.08t$, the intensity of the peak at $(0.35\pi,0.35\pi)$ is nearly the same as that near $(\pi,\pi)$. Above $0.08t$, the former decreases and the latter increases, but the opposite case occurs for $\omega \le 0.08t$. Secondly, the damping rate of quasiparticles should be 
lower than $\Delta_{0}$ due to the same reason. Numerical result shown in Fig.3(d) supports this analysis.
As for the case of nodeless $s$-wave symmetry as suggested by recent experiments~\cite{els}, we find that the spin response is suppressed to be only 5\% of that for $d_{x^{2}-y^{2}}$ wave and $\omega=0.02t$, due to the fully opened gap, and its structure is quite similar to that in the normal state shown in Fig.2. Therefore, the difference in the positions of IC peaks can be taken to differentiate the symmetry of the pairing states thus far proposed for $\kappa$-(ET).

Due to the underlying anisotropic triangular lattice in $\kappa$-(BT)~\cite{rev}, its SC gap symmetry is not a pure $d$ wave. We mimic the possible variation of the SC gap nodes by the mixed $d$ and isotropic $s$ wave. This is reasonable,
because the essential point we suggested in this paper is the position of the nodes, instead of the gap shape.  
In the case of $\Delta_{k}=c\Delta_{0}(\cos k_{x}-\cos k_{y})/2+(1-c)\Delta_{0}$, the nodes at $(k_{0},k_{0})$ and $(-k_{0},-k_{0})$  shift clockwise. Therefore the IC peak which is 
along the diagonal direction for pure $d_{x^{2}-y^{2}}$ wave also shifts clockwise. This can be seen clearly from Fig.4(a) with $c=0.8$. Because the shape of Fermi surface is symmetric with the axis along the diagonal direction, the $\pm 0.2\Delta_{0}$ terms will have a symmetry effect. However, the situation is different for a mixed $d_{xy}$ wave with a small $s$-wave component, due to the lack of the symmetry of Fermi surface with the axis $k_{x}$ or $k_{y}$ which is the node direction of a pure $d_{xy}$ wave. For a small positive $s$-wave component, the node at $(p_{1},0)$ will shift downwards from the $k_{x}$ axis to the position being more near the nesting wavevector $(0.35\pi,0.35\pi)$. Therefore, the IC peak at
$(0.35\pi,0.35\pi)$ is enhanced and has an much more overwhelming intensity than those peaks around $(\pi,\pi)$,
as shown in Fig.4(b). But for a negative $s$-wave component, the node at $(p_{1},0)$ will shift oppositely, i.e., move away from the nesting wavevector, and the peak at $(0.35\pi,0.35\pi)$ is strongly suppressed(Fig.4(c)). However, this mixed gap symmetry can still be distinguished from the pure or mixed $d_{x^{2}-y^{2}}$ wave. Because four peaks around $(\pi,\pi)$, instead of two peaks as in the pure or mixing $d_{x^{2}-y^{2}}$ symmetry, should be observed as a result of the suppression of the peaks along the diagonal direction where the gap is fully opened now.  

Finally, we discuss the $\omega$-dependence of the spin response, which is shown in Fig.5 for (a) $d_{x^{2}-y^{2}}$ wave and
(b) $d_{xy}$ wave, respectively. One can see that no resonance peak is found at momentum $q=(\pi,\pi)$(also for 
$q=(0.35\pi,0.35\pi)$ in the case of $d_{xy}$ wave). In the particle-hole channel, a spin resonance is expected if we have $1-\bar{U}{\rm Re}\chi _{0}({\bf q},\omega )=0$ and negligibly small Im$\chi _{0}({\bf q},\omega )$~\cite{li,blu}. 
This satisfies in high-$T_{c}$ cuprates~\cite{li,blu}, due to a sharp step-like rise of the bare Im$\chi _{0}$ at its gap edge, which comes from the extended van Hove singularity near $(0,\pi)$, then a logarithmic singularity in Re$\chi _{0}$ via
the Kramers-Kroenig relation. However, it does not satisfy here as see from
the inset of both figures 5(a) and (b), where the largest value of Re$\chi _{0}$ is about 0.249 and 0.262, respectively. The reason is that the step-like rise is very small or lack in Im$\chi _{0}$, because of the different energy band structure in $\kappa$-(BT). So, no resonance peak can be seen even for the largest value $\bar{U}=3.4t$ as shown in Fig.5(a).
The only condition for the occurence of a resonance peak here is limited to be $\bar{U}=4.0t \pm 0.1t$ for $d_{x^{2}-y^{2}}$ wave at $(\pi,\pi)$, but it is beyond the reasonable parameter range as discussed above. The above result is in sharp contrast with the $\omega$-dependence of the spin responses in high-$T_{c}$ cuprates and should have
important insight on the differences of the superconducting properties in both materials~\cite{li}.    

In summary, we propose an orientational-dependent method to determine the SC pairing symmetry in $\kappa$-(BEDT-TTF)$_{2}$X based on the $q$-space position of the incommensurate peaks of the spin susceptibility. It is demostrated that the incommensurate peak positions at low energies are uniquely determined by the pairing symmetry and the geometry of Fermi surface. Different patterns of the spin responses in the momentum space are shown for $d_{x^{2}-y^{2}}$ wave, $d_{xy}$ wave, mixed $d_{x^{2}-y^{2}}$ wave and mixed $d_{xy}$ wave. In addition, we predict the nonexistence of the spin resonance mode in this material.

Enlightening discussions with W.Y. Zhang, T. Xiang, X. Q. Wang and Z.B.Su are acknowledged. The project was 
supported by the National Nature Science Foundation of China and the Minstry of Science and Technology of China(NKBRSF-G19990646).

\newpage
\section*{FIGURE CAPTIONS}

Fig.1  Fermi surface in the first and the extended Brillouin zone. The arrows show particle-hole excitations with nesting wavevectors $(0.9\pi,0.9\pi)$(labeled as 1) and $(0.35\pi,0.35\pi)$(2), respectively.

Fig.2 Momemtum dependence of Im$\chi ({\bf q},\omega)$ in the normal state (temperature $T=0.01t$) for (a) $\omega=0.02t$ and (b) $\omega=0.2t$.

Fig.3 Momemtum dependence of Im$\chi ({\bf q},\omega)$ in the superconducting state (temperature $T=0.001t$) with $\omega=0.02t$ for (a) $d_{x^{2}-y^{2}}$ and (b) $d_{xy}$. (c) and (d) are the results along the diagonal direction with $d_{xy}$ for different excitation energies and different quasiparticle damping rates, respectively.

Fig.4 Momemtum dependence of Im$\chi ({\bf q},\omega)$ in the superconducting state (temperature $T=0.001t$) with $\omega=0.02t$ for (a) $0.8d_{x^{2}-y^{2}}+0.2$ isotropic $s$ wave, (b) $d_{xy}+0.2$ isotropic $s$ wave, and (c) $d_{xy}-0.2$ isotropic $s$ wave.

Fig.5 Frequency dependence of Im$\chi ({\bf q},\omega)$ for (a) $d_{x^{2}-y^{2}}$ and (b) $d_{xy}$ waves.
The insets show the frequency dependence of the bare spin susceptibility Im$\chi_{0} ({\bf q},\omega)$. Solid line
is their real part and dashed line their imaginary part. We have added the imaginary part a constant 0.23
and 0.22 for $d_{x^{2}-y^{2}}$ and  $d_{xy}$, respectively, in order to plot it in the same figure with their real part.

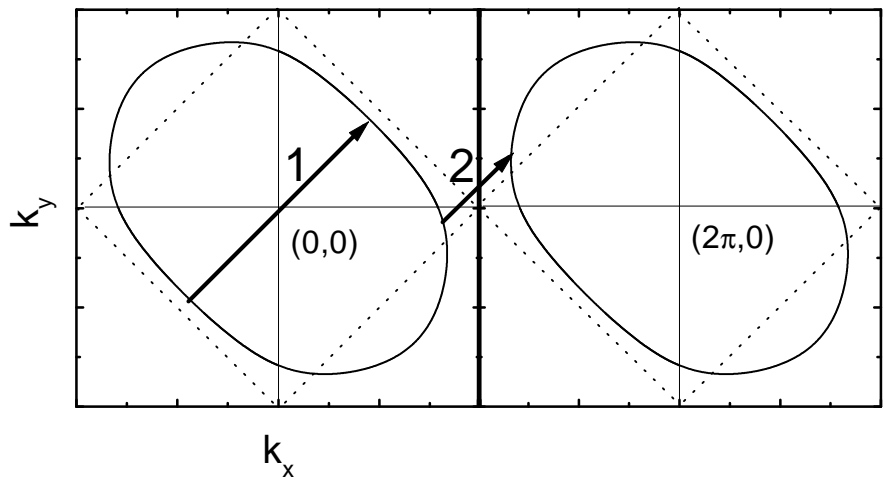
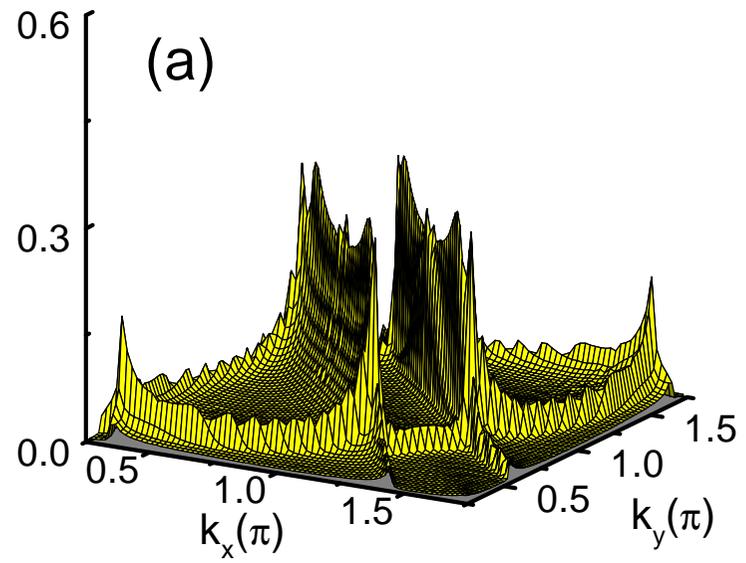
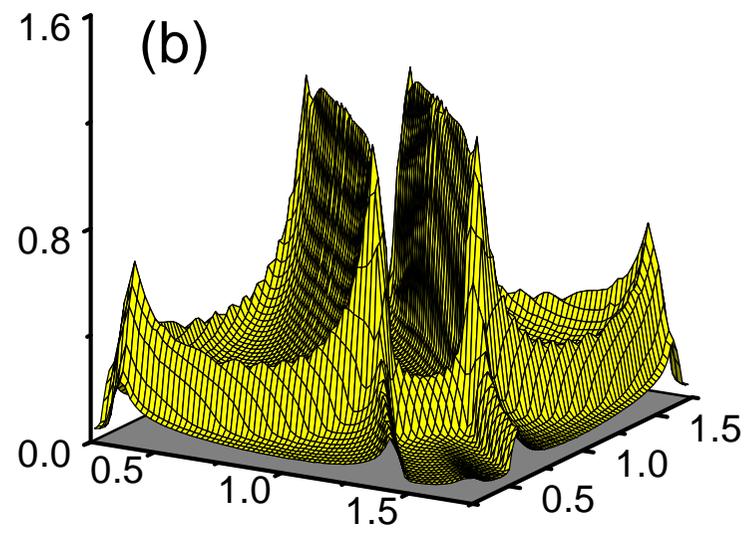
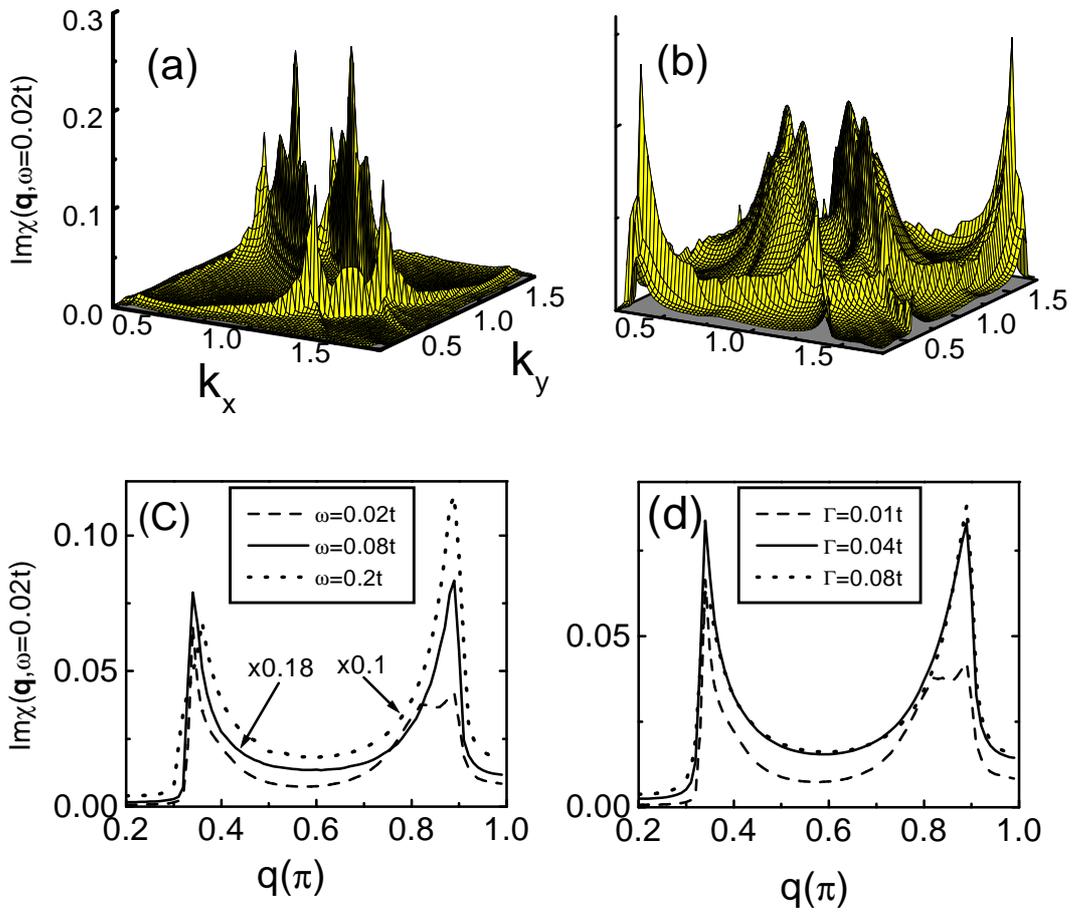
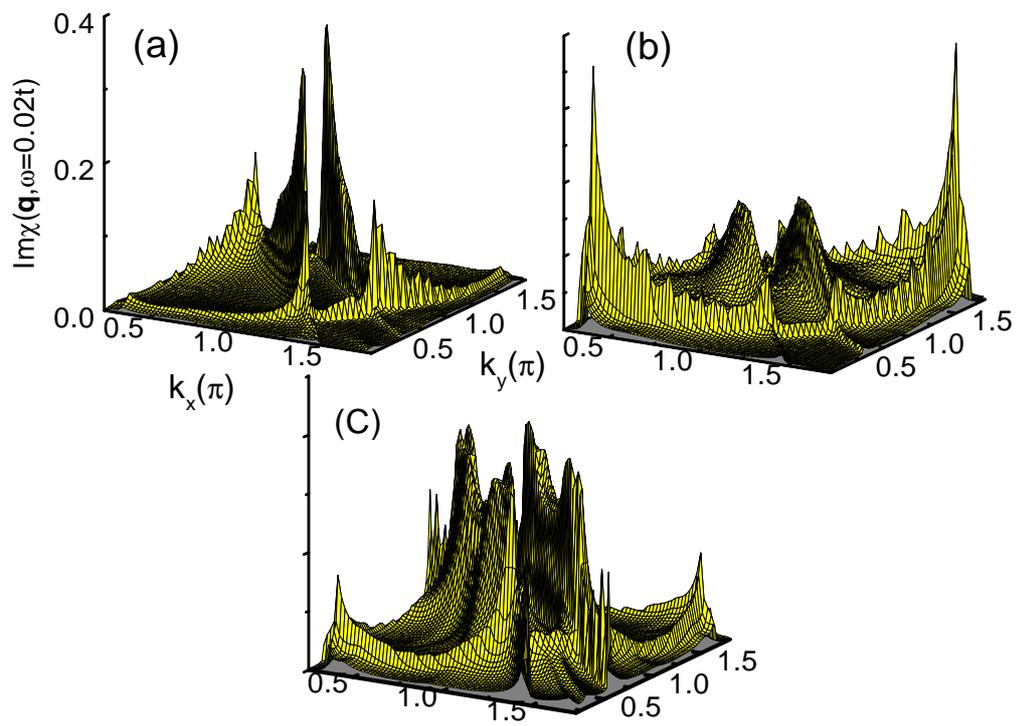
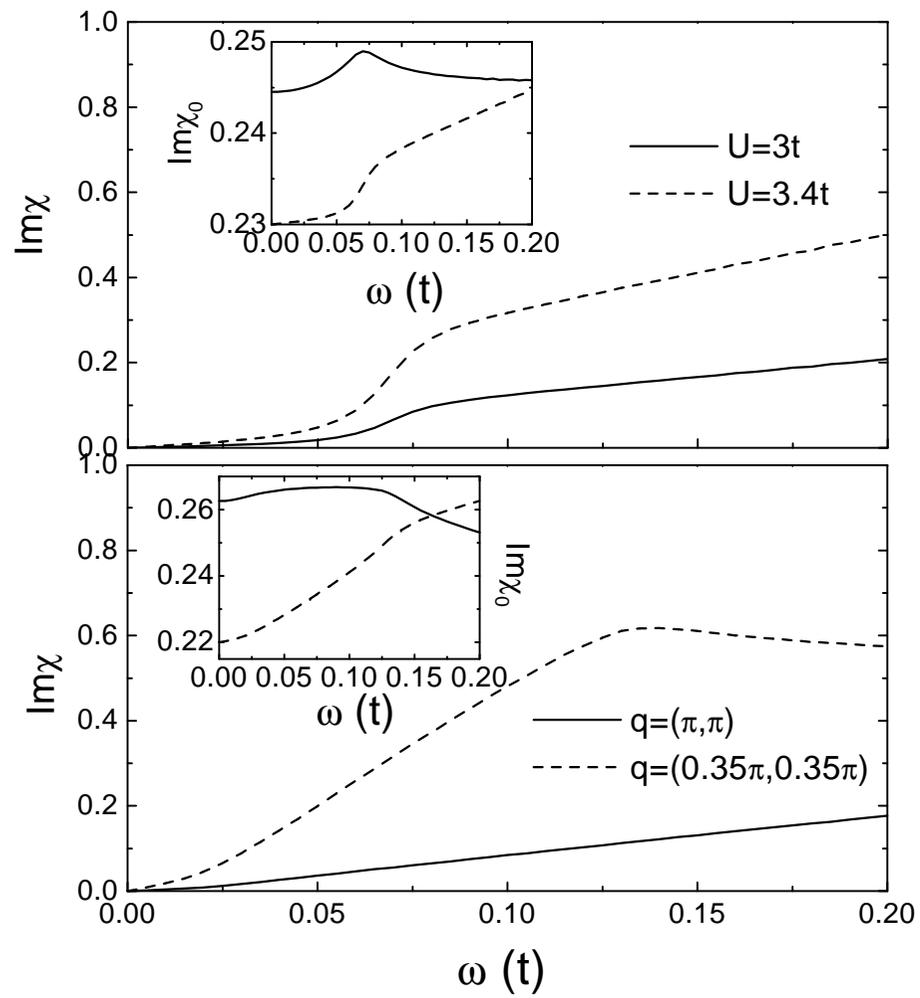
\end{document}